\documentclass[aps,prl,superscriptaddress,reprint ]{revtex4-1}%
\usepackage{epsfig,pslatex,latexsym,times,amssymb,amsmath,graphicx}
\usepackage{bm, float, lipsum}
\usepackage{amsmath}
\usepackage{amsfonts}
\usepackage{amssymb}
\usepackage{mathrsfs}
\usepackage{color, endnotes}
\usepackage{graphicx}%

\usepackage[normalem]{ulem}

\providecommand{\U}[1]{\protect\rule{.1in}{.1in}}
\begin{document}
\author{N. J. Harmon}
\email{nharmon@coastal.edu} 
\affiliation{Department of Physics and Engineering Science, Coastal Carolina University, Conway, SC
29526, USA}
\author{E. Z. Kurth}
\affiliation{Department of Physics and Engineering Science, Coastal Carolina University, Conway, SC
29526, USA}
\author{D. Coleman}
\affiliation{Department of Physics and Engineering Science, Coastal Carolina University, Conway, SC
29526, USA}
\author{L. Flanigan}
\affiliation{Department of Physics and Engineering Science, Coastal Carolina University, Conway, SC
29526, USA}

\date{\today}
\title{Generation of charge current by the Inverse Stern-Gerlach Effect and the suppression of spin transport due to spin counter-current exchange in semiconductors}
\begin{abstract}
The spin-orbit interaction is frequently the mechanism by which spin and charge are coupled for spintronic applications. The discovery of spin, a century ago, relied on spin-charge coupling by a magnetic field gradient; this mechanism has received scant attention as a means for generating spin and charge currents in semiconductors. Through the derivation of a set of coupled spin-charge drift-diffusion equations, our work shows that magnetic field gradients can be used to generate charge currents from non-equilibrium spin polarization, in solid state systems. We predict, in GaAs, an ``Stern-Gerlach" voltage comparable to what is measured by the inverse spin Hall effect. Non-intuitively, we find the spin diffusion length is reduced by the magnetic gradient. This is understood by invoking the idea of co-current and counter-current exchange which is a concept frequently invoked in fields as disparate as animal physiology and thermal engineering.
\end{abstract}
\maketitle	

\emph{Introduction. --- }
In 1922, Otto Stern and Walther Gerlach discovered quantized spin angular momentum in neutral silver atoms when an inhomogeneous magnetic field spatially separated discrete spin states \cite{gerlach1922a, gerlach1922b,Fraser1927}. The Stern-Gerlach experiment's detection of spin was indirect as it measured the spin of a neutral atom and not the spin of a free charge. Debate arose in the early days of quantum theory whether the spin of a \emph{free} electron could be verified with a Stern-Gerlach-like experiment \cite{brillouin1927, brillouin1928, Garraway2002}. The conventional wisdom that emerged (though detractors exist \cite{Garraway1999, Gallup2001}), repeating the arguments of N. Bohr and N. F. Mott \cite{Mott1929, Kessler1985, fabian2007}, is that Lorentz forces obscure any magnetic moment deflections. Due to $\bold{\nabla} \cdot \bold{B} = 0$, the Lorentz force cannot be avoided by directing motion of spin  carriers parallel to $\bold{B}$ since a gradient field force transverse to the motion also exists. 

What is overlooked, aside from a few authors \cite{Fabian2002,Martin2003, Wrobel2004}, is these arguments apply for \emph{free} electrons which is not the condition which electrons exist in solid state systems. In a confined geometry, the Lorentz force, which is the primary obstacle to spin separation, and the transverse gradient force (due to $\bold{\nabla} \cdot \bold{B} = 0$), would be negated by the classical Hall effect. 
Fabian and Das Sarma studied the effect of inhomogeneous fields on one dimensional conduction spin currents by solving the Boltzmann equation for a limited set of boundary conditions \cite{Fabian2002b}. Their calculations suggest the spin current generated is constrained by the time for a carrier to diffuse across the sample and the spin relaxation time. Our focus here is on the generation of  more experimentally accessible charge currents.

The prospect of spin-charge coupling, outside of the spin-orbit interaction \cite{Dyakonov1971, Dyakonov1971a, nagaosa2010, Sinova2015, Edelstein1990, Ganichev2002, Kato2004c}, is alluring as it offers up a wider variety of materials for inspection. What will be demonstrated in this Letter, by formulating and solving a series of spin and charge drift-diffusion equations, is that magnetic field gradients induce charge currents and spin accumulation which are predicted to be measurable by a voltage difference. This is not too surprising since one can view the original Stern-Gerlach experiment as an unpolarized beam generating a transverse spin current; here we have a spin polarization generating a charge current -- a process we term as the Inverse Stern-Gerlach effect. However the role the field gradient, in conjunction with spin relaxation, plays in suppressing spin diffusion is subtle and is best understood by analogy with the concept of co- and counter-current exchange from the fields of physiology \cite{vandam1938, Hazelhoff1952, Hughes1973, Ullrich1961} and fluid/thermal dynamics \cite{Randall1939, Schicks2019}.

\begin{figure}[H]
 \begin{centering}
        \includegraphics[scale = 0.3,trim = 18 40 150 10, angle = -0,clip]{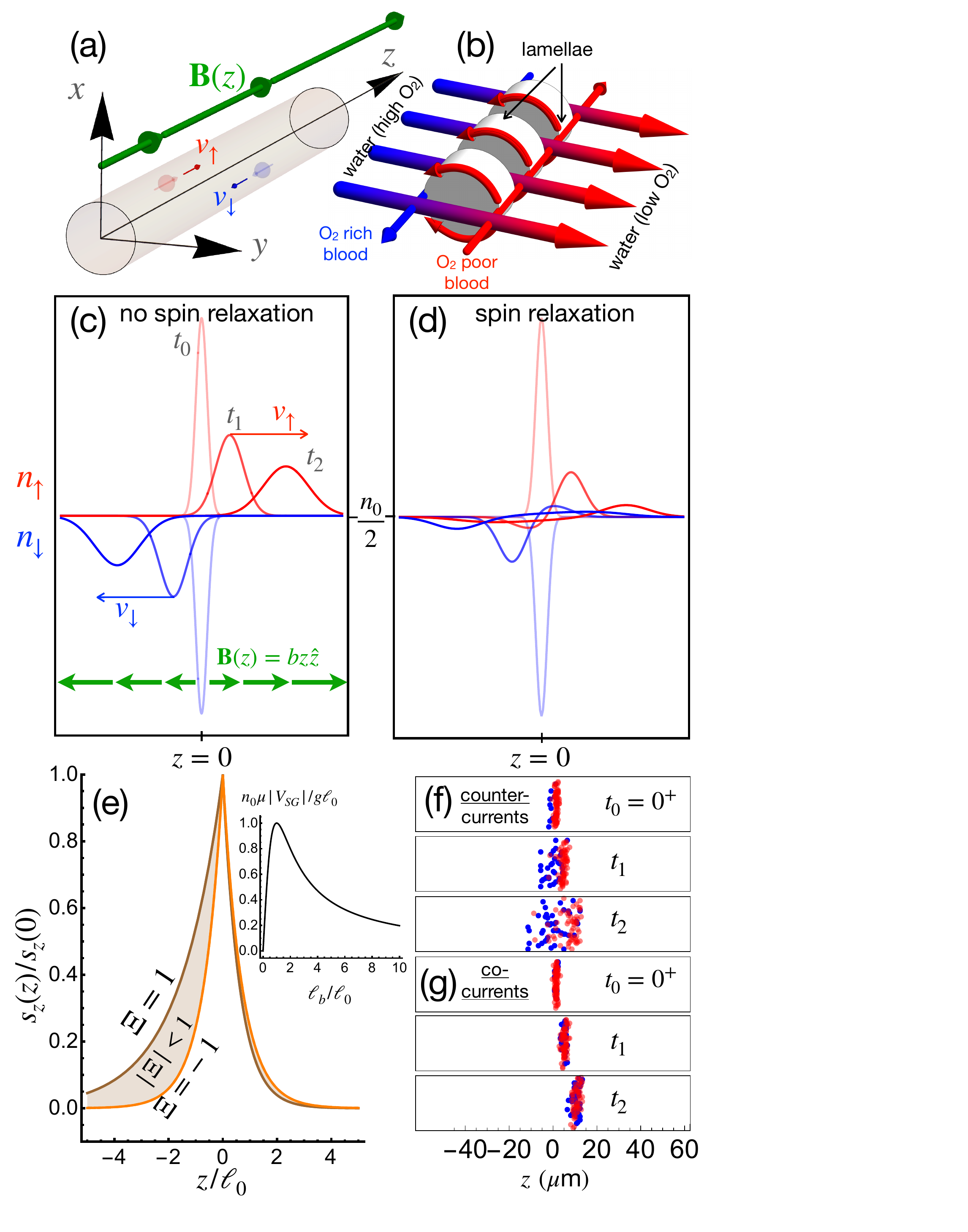}
        \caption[]
{(a) Geometry of model quasi-1d model system. Magnetic field, $\bold{B}(z)$, and spin lie along $z$ axis. (b) Example of counter-current exchange of CO$_2$ and O$_2$ in fish gills. (c) Up and down spin densities for three different times after an initial gaussian spin injection at $z = 0$.  The spin injection leads to deviations of up and down densities away from their equilibrium value $n_0/2$. Under the influence of the magnetic field gradient force and diffusion, up and down spin densities flow in opposite directions. 
Inset: the non-uniform magnetic field $\bm{B}(z)$ considered in this work. 
(d) Identical to (c) except spin relaxation is included.  Direction of packet motion is determined by $\nu b$ which is chosen here as $\nu b<0$. 
(e) Allowable steady state normalized spin density region for possible spin cc exchange parameters, $\Xi$, for $\tau_s' = D' = 1.$ The brown (orange) line denotes spin co-current (counter-current) exchange given by Eqs. \eqref{eq:dimless1} and \eqref{eq:dimless2}. The influence of counter-current exchange is to reduce the (longer) downstream spin diffusion length and slightly increase the (much shorter) upstream spin diffusion length. Inset: predicted dependence of Stern-Gerlach voltage, $V_{SG}$, on magnetic gradient length, $\ell_b$.
(f,g) Results, at three different instances, from Monte Carlo simulation of up spins (red) injected into quasi-one-dimensional channel under condition of (f) spin counter-current exchange ($\nu b \neq 0$, $E_z = 0$) (g) and spin co-current exchange ($\nu b = 0$, $E_z \neq 0$). Due to spin relaxation, up spins convert to down spins (blue) and vice-versa.
Ordinate quantities in (c,d) are per $\delta N_{\uparrow}/A$.}\label{fig:nupdown} 
        \end{centering}
\end{figure}

\begin{figure}[h]
 \begin{centering}
        \includegraphics[scale = 0.42,trim = 180 375 120 10, angle = -0,clip]{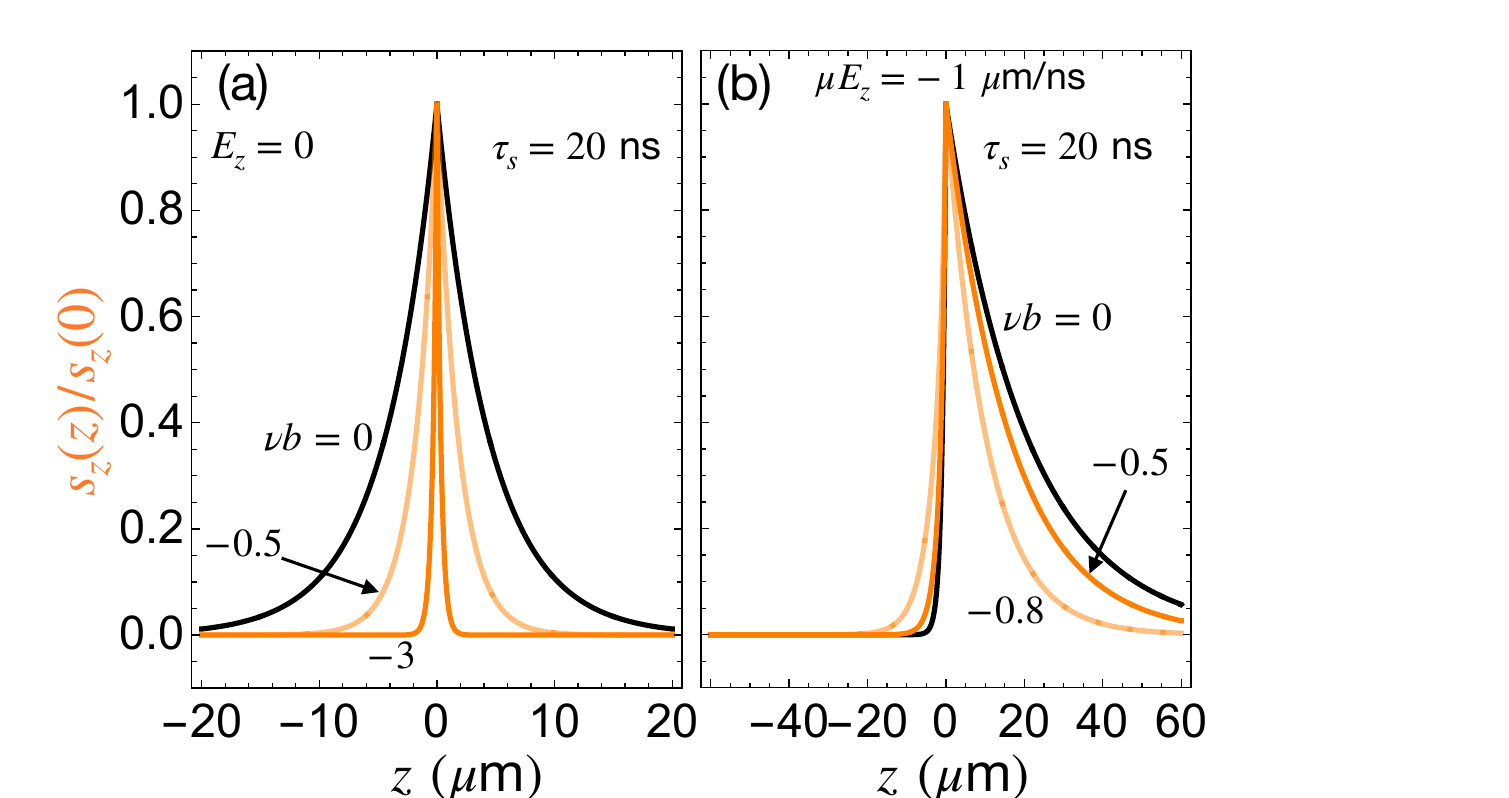}
        \caption[]
{Calculations of steady state spin density (a) and up/down spin densities (b) in the presence of the constant magnetic field gradient in the absence (a,b) and presence (c) of a uniform electric field. A delta source of spin injection at the origin is assumed. (a,b) are normalized to clearly demonstrate how $b$ influences spin diffusion length. Ordinate quantities in (c) are per $g$.}\label{fig:steadystate} 
        \end{centering}
\end{figure}

Our main results, for a one-dimensional system (Fig. \ref{fig:nupdown}(a)), are shown in Fig. \ref{fig:nupdown}. Fig. \ref{fig:nupdown}(c) displays the transient responses of up/down spin densities after a gaussian impulse of spin density in the absence of spin relaxation; up and down spin travel in opposite directions \emph{leading to a charge current} which, due to the nature of the magnetic-gradient interaction, depends upon the spin density instead of the charge density. Fig. \ref{fig:nupdown}(d) plots the same quantities which demonstrate new features (anti-nodes) when spin relaxation is present.  A 
Fig. \ref{fig:steadystate}(e) depicts steady state spin distributions for different field gradients (brown line -- small gradient, orange line -- large gradient). Despite the Stern Gerlach's determination to separate opposite spins in opposite, the spin diffusion lengths \emph{shrink} as a result; we seek to understand this phenomena by analogy with counter-current exchange (an example of counter-current exchange in fish oxygenation is shown in Fig. \ref{fig:steadystate}(b)).
The charge separation of Figs. \ref{fig:steadystate}(c,d) produces a gradient-dependent voltage -- a predicted response inset in Fig. \ref{fig:nupdown}(e).

\emph{Spin and charge drift-diffusion equations in presence of magnetic field gradient. --- }
We approach the problem in a semi-classical manner by solving for spin-dependent velocities resulting from a Stern-Gerlach-like force from a generic magnetic field gradient.
Given the magnetic moment of an electron, $\bm{\mu}  = -\frac{1}{2} g \mu_B \bm{\sigma}$ and the its dipole potential energy, $V =  - \bm{\mu} \cdot \bm{B}$, we write a Stern-Gerlach force operator as
\begin{equation}
 \hat{\bm{F}}_{SG} = - \nabla V = - \frac{1}{2} g \mu_B  \nabla (\bm{\sigma}\cdot \bm{B} ).
\end{equation}
The force is spin-dependent --- the force on $\uparrow/\downarrow$ spins (taking $z$ as quantization axis) is $\bm{F}_{SG}^{\uparrow/\downarrow} =  \mp  \frac{1}{2} g \mu_B  \nabla B_{z}$.
The inclusion of this force into a Drude-like model yields steady state velocities 
\begin{equation}
\bm{v}^{\uparrow}  = \frac{\bm{F}_{SG}^{\uparrow} \tau}{m} = - \frac{g \mu_B \tau}{2m} \nabla B_z =  -\nu \nabla B_z,
\end{equation}
where $\nu =\frac{g^* \mu_B \tau}{2m}  $ has units of m$^2$/(T~s) which we define as the \emph{magnetic mobility} (in contrast to electronic mobility), $\tau$ is the momentum relaxation time, $m$ is the carrier mass, and $g^*$ is the effective $g$-factor. 
Note that the velocity does not depend on the charge $q$ in congruence with the force being moment- and not charge-dependent.
The spin current is $\bm{j}_{SG,s}^{z}  = \bm{j}_{SG}^{\uparrow} -  \bm{j}_{SG}^{\downarrow} = - q   (n_{\uparrow} + n_{\downarrow}) \nu   \nabla B_{z} = - q   n \nu   \nabla B_{z}$ where $n$ is the conduction charge density. Similarly for other directions of spin, the direction of spin couples to the same direction of magnetic field: $\bm{j}_{SG,s}^{i} = - q   n \nu   \nabla B_{i}$.
We follow the same process for charge current:
$\bm{j}_{SG,c} =  -  q  \nu (   s_x \nabla B_{x} +  s_y \nabla B_{y} +  s_z \nabla B_{z})$ where $s_i$ are components of spin density.

Following the usual prescription \cite{kos2007, Wu2010, Harmon2015, obrien2016}, the continuity equations for charge and spin are as follows:
\begin{equation}
\frac{\partial n}{\partial t} = -\frac{1}{q} \nabla \cdot \bm{j}_c, ~~
\frac{\partial s_{i}}{\partial t} = -\frac{1}{q} \nabla \cdot \bm{j}_s^{i}   -  \frac{s_{i}}{\tau_s} +  \text{sign}(q) \frac{g\mu_B}{\hbar} [\bm{s} \times \bm{B}(\bm{r})]^{i}
\end{equation}
in which we substitute in each type of current: diffusion ($\bm{j}_{diff}  = -q D \nabla n$), drift ($\bm{j}_{drift}  = \text{sign}(q) q n  \mu  \bm{E}$), and Stern-Gerlach (as just derived) where $\mu = |q| \tau/m$ is the charge mobility, $D$ is the diffusion constant, and $\tau_s$ is the spin relaxation time. Spin current versions for diffusion and drift are found simply by $n \rightarrow s_i$. After insertion of currents into the continuity equations we obtain a set of four coupled drift-diffusion equations for charge and spin:

\begin{eqnarray}
 \frac{\partial n}{\partial t} &=&  D \nabla^2 n - \text{sign}(q) \mu [n \nabla \cdot \bm{E}  + (\nabla n )\cdot \bm{E}] + {}\nonumber \\ 
 {}&+&\nu \sum_{i \in \{x,y,z\}} \left[ (\nabla s_i) \cdot \nabla B_{i}(\bm{r})  + s_i \nabla^2 B_{i} (\bm{r})
\right] \label{eq:chargeDD}   
\end{eqnarray}
\begin{eqnarray}
\frac{\partial s_i}{\partial t} &=&  D \nabla^2 s_i  -  \text{sign}(q) \mu [ s_i \nabla \cdot \bm{E}  +  (\nabla s_i )\cdot \bm{E}] - \frac{s_i}{\tau_s}  + {}  \\
{} &+& \nu \left[ (\nabla n) \cdot \nabla B_{i}(\bm{r})  + n \nabla^2 B_{i} (\bm{r})
\right] +  \text{sign}(q) |\gamma| \left(  \bm{s} \times \bm{B}(\bm{r}) \right)_i.\nonumber
\label{eq:spinDD}
\end{eqnarray}

To understand the physical ramifications of the field gradient, in this Letter we focus on the simplest case of one dimension ($z$) with electron carriers in a uniform electric field, an external magnetic field pointing along the one dimension such that $\bm{B}(z) = (B_0 + b z )\hat{z} $ (which avoids any complications of Lorentz force and spin precession) in order to focus solely on the role of the gradient term $bz$, and any injected spin is oriented along $z$ as shown in Fig. \ref{fig:nupdown}(a). With these simplifications, equations \eqref{eq:chargeDD} and \eqref{eq:spinDD} reduce to
\begin{equation}
    \frac{\partial n}{\partial t} =  D \partial_z^2 n  + \mu E_z \partial_z n +  \nu b \partial_z s_z 
    \label{eq:spde}
\end{equation}
\begin{equation}
    \frac{\partial s_z}{\partial t} =  D \partial_z^2 s_z + \mu E_z \partial_z s_z + \nu b \partial_z n- \frac{s_z}{\tau_s}.
    \label{eq:npde}
\end{equation}
Suggesting a specific experimental realization determines our boundary and initial conditions.

\emph{Transient spin dynamics. --- }
In an $n$-doped semiconductor, carriers can be spin polarized through methods of optical orientation \cite{Meier1984}. The process is briefly summarized as circularly polarized light exciting partially spin-polarized electrons, $n_{ex}$,  into the conduction band which has an equilibrium concentration $n_0 \gg n_{ex}$.
Conduction electrons rapidly recombine with conduction holes on a time scale $\tau_r<< \tau_s$. Since there are more unpolarized electrons than polarized electrons, mostly unpolarized electrons recombine which leaves a net spin density in the conduction band while the concentration of conduction electrons has returned to equilibrium, $n(0,z) = n_0$, after recombination \cite{Kikkawa1998}. 
It is after recombination, while a finite conduction spin density ($s_z(0,z) = 2\frac{S_0}{A}\delta(z)$) remains, that we consider our starting point as we solve Eqs. \eqref{eq:spde} and \eqref{eq:npde}.
$2S_0$ is the initial difference in spin up ($N_{\uparrow}$) and spin down ($N_{\downarrow}$) electrons and $A$ is the cross sectional area of the quasi-one-dimensional system \cite{supp}.

\begin{figure}[hbt]
 \begin{centering}
        \includegraphics[scale = 0.9,trim = 1 1 550 80, angle = -0,clip]{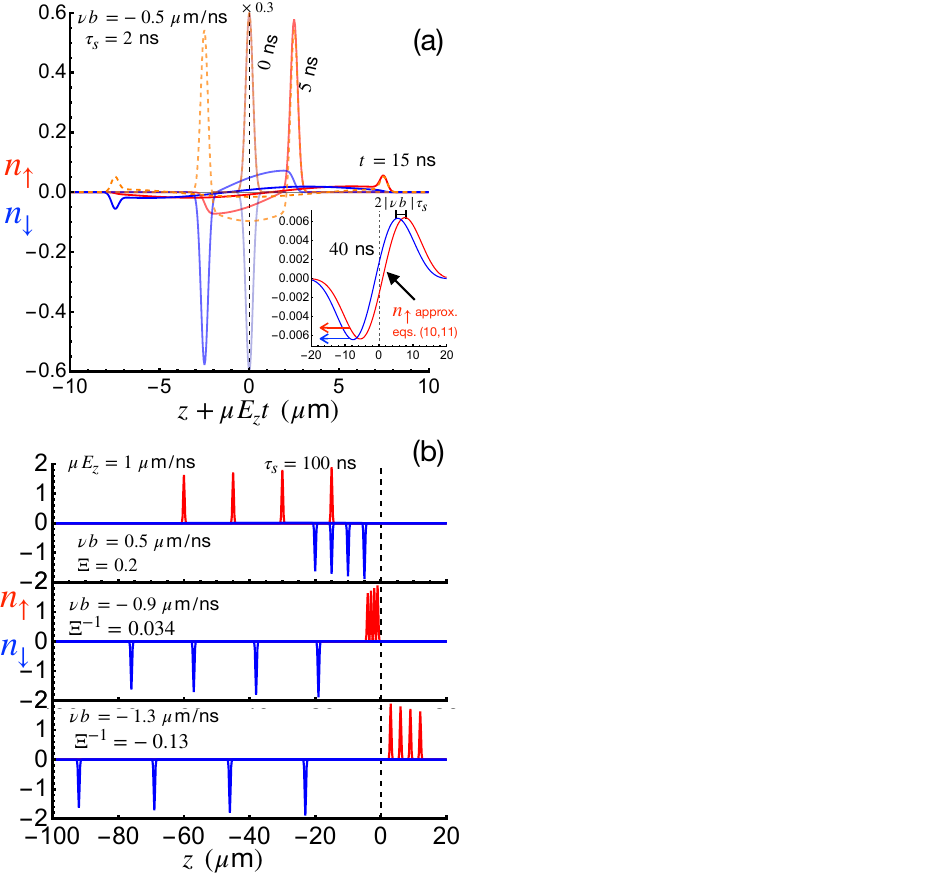}
        \caption[]
{(a)  Examples of spin carrier motion (up-red, down-blue, spin-orange) for moderate spin relaxation at four different times. The initial gaussian peaks moves ballistically but is reduced in size quickly due to spin relaxation. A slowly decaying, broad, shallow peak follows in its wake as spins flip back and forth. The inset shows the up and down spins at longer time when the narrow peak has vanished, leaving behind the broad diffusive peaks that maintain a separation due to the opposite Stern Gerlach forces on them.
The $t = 0$ traces are downsized by a factor of 0.3. 
(b) Examples of spin carrier motion for different spin cc exchange parameters, $\Xi$, when spin relaxation is weak; up and down spin travel and different speeds depending up $E_z$ and $\nu b$. Snapshots are at times 10, 20, 30, and 40 ns. For all, $\sigma = 0.2$ $\mu$m and $\delta$ impulse. 
Ordinate quantities are per $\delta N_{\uparrow}/A$ where $\delta N_{\uparrow} = N_{\uparrow} - N_0/2$.}\label{fig:transientWspin} 
        \end{centering}
\end{figure}

It is instructive to examine the  time-dependent solutions to eqs. \eqref{eq:spde}, \eqref{eq:npde} in terms of $n_{\uparrow,\downarrow}$. The coupling between $n_{\uparrow}$ and $n_{\downarrow}$ comes from spin-flip processes alone \cite{supp}.
Fig. \ref{fig:nupdown}(d) shows the effects of spin relaxation: peaks are reduced as expected but also additional broad peaks form. This is further seen in Fig. \ref{fig:transientWspin}(a) where a narrow peak quickly drops off while a much broader trailing peak rapidly forms and then slowly decays in time. At long times (Fig. \ref{fig:transientWspin}(a) inset), each $n_{\uparrow}$ and $n_{\downarrow}$ are approximately odd functions with peaks that slowly continue to lessen. The peaks on either side of the origin move in opposite directions \emph{for both $n_{\uparrow}$ and $n_{\downarrow}$} in seeming contradiction to the Stern Gerlach forces. The cause of this behavior can be understood by considering two up spins moving to right. One flips and begins moving to left (via Stern Gerlach), separating from its partner. It may then flip again, causing it to move right but still trailing its partner. This process continues with the net result being what we observe in Fig. \ref{fig:transientWspin}(a) inset. We can find quantitative results by successive Fourier and Laplace transforms on eqs. \eqref{eq:spde}, \eqref{eq:npde} if $D$ and impulse width are set to zero (which do not change the pertinent results). In the long-time limit, 
\begin{equation}\label{eq:longs}
    s_z =  \frac{  e^{-\frac{\gamma_s (z+\mu E_z t)^2}{4 (\nu b)^2  t}} }{2 A \sqrt{\pi} |\nu b| \sqrt{\gamma_s}~t^{3/2}} \delta N_{\uparrow} \left( \frac{\gamma_s (z+\mu E_z t)^2}{2|\nu b|^2 t} - 1 \right)
\end{equation}
and
\begin{equation}\label{eq:longn}
   \delta n =  -\text{sign}(\nu b)\frac{ \sqrt{\gamma_s} (z+\mu E_z t) e^{-\frac{\gamma_s (z+\mu E_z t)^2}{4 (\nu b)^2  t}} }{2 A \sqrt{\pi} |\nu b|^2 ~t^{3/2}} \delta N_{\uparrow}.
\end{equation}
From these results we find the separation between the peaks $\sim 2 |\nu b| \tau_s$ --- if spin relaxation is very strong, flipped spins do not have time to backtrack so the peaks coincide and the spin density is zero. The broad pulses also exist for $s_z$ (dashed, orange); their peak positions are $z_s = \pm \sqrt{6 t \tau_s}|\nu b |$ which suggests spin relaxation induces diffusive-like transport as spin flip between the up and down channels.

When an electric field is present, both spin species experience the same electric force which acts essentially like a change of reference frame for the spin packets. 
Fig. \ref{fig:transientWspin}(b) displays three transients where both the gradient force and electric force exist. Depending on the relative sizes of the two forces, the up and down spins move parallel or anti-parallel to one another.
We relate these results to the concepts of co/counter-current (cc) exchange in which bipartite flows are either parallel (co) or anti-parallel (counter). 
A canonical example of cc exchange is fish respiration where O$_2$ rich water flows between gill filament lamellae as shown in Fig. \ref{fig:nupdown}(b). Gas exchange occurs in the lamellae as CO$_2$ in blood is exchanged for O$_2$ in the water. In this manner, oxygen enters the bloodstream. The exchange is efficient because the blood flows opposite that of water. Intuitively, counter-current exchange can be thought of as more efficient than co-current exchange since to increase oxygen, deficient blood should move toward the source of oxygen (counter-current) and not away (co-current). The flow of up and down spins is analogous where the exchange process is spin flips between the up and down spin channels.
The diffusive behavior determined above in Fig. \ref{fig:transientWspin}(a) inset (counter-current since $b\neq 0$ and $E_z = 0$) is not present when $b = 0$ and $E_z \neq 0$ (co-current) where purely exponential in time behavior arises \cite{supp}. 

\emph{Steady state. --- }
Alternatively, we consider continuous optical pumping of spin into the conduction band \cite{Meier1984, Kikkawa1997} to determine steady state spin and charge concentrations \cite{supp}.
If $g$ is the rate per cross-sectional area of continuously injected non-equilibrium spins at the origin, the solutions can be written in the following compact form
\begin{equation}
    s_{z}=g \tau_s \frac{e^{-\frac{|z|}{\ell_0^2}\sqrt{(\ell_b^2+ \ell_0^2)}}}{\sqrt{\ell_b^2+ \ell_0^2}}, 
    \label{eq:steadystateSZ}
\end{equation}
\begin{equation}
    \delta n= g\tau_s \text{sign}(\nu b) \ell_b  \frac{e^{-\frac{|z|}{\ell_0^2}\sqrt{\ell_b^2+ \ell_0^2}}-1}{\ell_b^2+ \ell_0^2}\text{sign}(z)
    \label{71}
\end{equation}
where $\ell_0 = \sqrt{D \tau_s}$ is the spin diffusion length and $\ell_b = |\nu b| \tau_s$ magnetic gradient drift length. 
When $\ell_b \gg \ell_0$, the diffusion lengths become $\ell_0^2/ \ell_b = D/|\nu b|
$ which explains the behavior for spin diffusion lengths seen in Fig. \ref{fig:steadystate} (a).
Though the spin-flip coupling between the two channels is the same, their opposite directions severely reduce the transport of spin (orange curves). This is spin counter-current exchange.
Fig. \ref{fig:steadystate}(b) shows a similar suppression of the downstream spin diffusion length when a electric field is applied. The electric field send each spin in same direction which reduces the severity of the counter-current exchange. 
Figs. \ref{fig:nupdown}(f,g) makes this apparent: the up and down spin populations separate from one another for pure counter-currents which means a down spin is unlikely to contribute to the primary packet when it flips back to up. For co-current exchange ($b = 0$), spin flips are less severe as up and down spins stay proximal and the integrity of the packet is maintained over a large distance (the downstream spin diffusion length).

To see the effects of cc exchange in general (for any $b$ and $E_z$) we rewrite the drift-diffusion equations for $n_{\uparrow, \downarrow}$ in dimensionless form:
\begin{equation}
    \frac{\partial n_{\uparrow}}{\partial {t'}} =  D' \partial_{z'}^2 n_{\uparrow}  +  \partial_{z'} n_{\uparrow} - \frac{n_{\uparrow} - n_{\uparrow}}{2 \tau_s'}
    \label{eq:dimless1}
\end{equation}
\begin{equation}
    \frac{\partial n_{\downarrow}}{{\partial {t'}}} =  D' \partial_{z'}^2 n_{\downarrow} + \Xi  \partial_{z'} n_{\downarrow} - \frac{n_{\downarrow} - n_{\uparrow}}{2 \tau_s'}
    \label{eq:dimless2}
\end{equation}
where $t' = t(\mu E_z + \nu b)/\ell_0$ and $D' = D/\ell_0(\mu E_z + \nu b)$, $\tau_s' =  \tau_s(\mu E_z + \nu b)/\ell_0$, and $z' = z/\ell_0$ are dimensionless units, $\Xi = \frac{\mu E_z - \nu b}{\mu E_z + \nu b}$ is the \emph{spin cc exchange parameter} which gives the degree of opposition present between up and down spin currents, and $\mu E_z$ and $\nu b$ are assumed to have same sign \cite{supp}. $\Xi = 1$ ($b = 0$) implies pure spin co-current exchange, and  $\Xi = -1$ ($E_z = 0$) implies pure spin counter-current exchange. Spin co-current exchange is simply what occurs in an electric field (no gradient term) and has been examined in detail by others \cite{Yu2002a, Yu2002, tahara2016}. When both electric field and gradient field are present an intermediate situation exists where $ -1<\Xi < 1$.
Fig. \ref{fig:nupdown}(e) shows any deviation from purely co-current conditions (black line) reduces the downstream spin diffusion length; the strong suppression occurs for purely counter-current exchange (orange line).

\emph{Stern Gerlach potential difference. --- }
Although the magnetic field gradient may be an obstacle for spin transport, it nevertheless does give rise to a Stern Gerlach charge current via $\bm{j}_{SG,c} =  -  q  \nu s_z \nabla B_{z}$.
The potential difference across the sample is found considering the Stern Gerlach force as an effective electric force such that $\bm{j}_{SG,c} =  \sigma_c \bm{E}_{\text{eff}}$ where $\sigma_c = |q|\mu n$ is the charge conductivity. We then write $
    \Delta V_{SG} =\int_{-\infty}^{\infty} \frac{j_{SG,c}}{\sigma_c} dz$.
With an electric field absent, in order to not obscure the Stern Gerlach potential, we use $s_z$ of eq. \eqref{eq:steadystateSZ} to obtain
\begin{equation}
    |\Delta V_{SG}| = \frac{2 g}{\mu n_0} \ell_b \frac{\ell_0^2}{\ell_0^2+\ell_b^2}
\end{equation}
where we use $n \approx n_0$ since $n_0 \gg \delta n$.  Fig. \ref{fig:nupdown}(e) inset demonstrates the relationship between $\Delta V_{SG}$ and the field gradient length. Maximum voltage is when $\ell_0 =\ell_b$; under such a condition, the voltage scales as $\ell_b$.

To estimate the potential we consider $g = s_{ex}f$ as the rate of generated spins per cross-sectional area. We approximate $s_{ex}\sim 10^9$ cm$^{-2}$ given a donor density of $\sim 10^{10}$ cm$^{-2}$ in a GaAs quantum well \cite{Ohno1999b} and $f\sim 100$ MHz \cite{Kikkawa1997, Ohno1999b}. A typical spin diffusion length in GaAs is 10 $\mu$m \cite{Lou2007, Furis2007}.
Using $\mu = 1000$ cm$^2$/Vs, the maximum voltage is $\sim 10$ $\mu$V which is larger than typical inverse spin Hall voltages in GaAs in the $\sim 1$ $\mu$V range \cite{olejnik2012}.
To achieve $\ell_b \sim 10$ $\mu$m, $b$ would need to be $\sim 10^6$ T/m which is large but comparable gradients have been produced by the stray fields of certain patterned magnetic films \cite{Cohen2009, kustov2010,zablotskii2016}. Rashba-field gradients offer even larger gradients $\sim 10^8$ T/m which have been used to produce spin separation in InAs quantum point contacts \cite{Kohda2012}
 Smaller, though still significant, are gradients produced by  spin polarized nuclei  \cite{Paget1977, Meier1984,Chan2009, Harmon2015,Harmon2022a}.
 Advantages to these latter two are avoidance of complications from Lorentz forces in two or three dimensions and, if appropriately oriented, effects of spin precession.

\emph{Conclusion.} --- The force of magnetic field gradients on conduction spin magnetic moments has been incorporated into a series of spin and charge drift-diffusion equations. We suggest an optical approach to measure an appreciable voltage difference generated by spin and charge dynamics in the magnetic gradient. The predicted Stern Gerlach voltage, in competition with an inverse Spin Hall voltage, may also be measured electrically in a two-dimensional geometry where a gradient field is transverse to a spin current. A reported dependence of a potential dependent on dynamically polarized nuclei is suggestive of this Inverse Stern Gerlach-like effect \cite{geppert2013}.

\begin{acknowledgments}
 This work was supported by the National Science Foundation under grant numbers DMR-2014786 and DMR-2152540.
\end{acknowledgments}


%











\begin{thebibliography}{51}%
\makeatletter
\providecommand \@ifxundefined [1]{%
 \@ifx{#1\undefined}
}%
\providecommand \@ifnum [1]{%
 \ifnum #1\expandafter \@firstoftwo
 \else \expandafter \@secondoftwo
 \fi
}%
\providecommand \@ifx [1]{%
 \ifx #1\expandafter \@firstoftwo
 \else \expandafter \@secondoftwo
 \fi
}%
\providecommand \natexlab [1]{#1}%
\providecommand \enquote  [1]{``#1''}%
\providecommand \bibnamefont  [1]{#1}%
\providecommand \bibfnamefont [1]{#1}%
\providecommand \citenamefont [1]{#1}%
\providecommand \href@noop [0]{\@secondoftwo}%
\providecommand \href [0]{\begingroup \@sanitize@url \@href}%
\providecommand \@href[1]{\@@startlink{#1}\@@href}%
\providecommand \@@href[1]{\endgroup#1\@@endlink}%
\providecommand \@sanitize@url [0]{\catcode `\\12\catcode `\$12\catcode
  `\&12\catcode `\#12\catcode `\^12\catcode `\_12\catcode `\%12\relax}%
\providecommand \@@startlink[1]{}%
\providecommand \@@endlink[0]{}%
\providecommand \url  [0]{\begingroup\@sanitize@url \@url }%
\providecommand \@url [1]{\endgroup\@href {#1}{\urlprefix }}%
\providecommand \urlprefix  [0]{URL }%
\providecommand \Eprint [0]{\href }%
\providecommand \doibase [0]{http://dx.doi.org/}%
\providecommand \selectlanguage [0]{\@gobble}%
\providecommand \bibinfo  [0]{\@secondoftwo}%
\providecommand \bibfield  [0]{\@secondoftwo}%
\providecommand \translation [1]{[#1]}%
\providecommand \BibitemOpen [0]{}%
\providecommand \bibitemStop [0]{}%
\providecommand \bibitemNoStop [0]{.\EOS\space}%
\providecommand \EOS [0]{\spacefactor3000\relax}%
\providecommand \BibitemShut  [1]{\csname bibitem#1\endcsname}%
\let\auto@bib@innerbib\@empty
\bibitem [{\citenamefont {Gerlach}\ and\ \citenamefont
  {Stern}(1922{\natexlab{a}})}]{gerlach1922a}%
  \BibitemOpen
  \bibfield  {author} {\bibinfo {author} {\bibfnamefont {W.}~\bibnamefont
  {Gerlach}}\ and\ \bibinfo {author} {\bibfnamefont {O.}~\bibnamefont
  {Stern}},\ }\href@noop {} {\bibfield  {journal} {\bibinfo  {journal}
  {Zeitschrift f{\"u}r Physik}\ }\textbf {\bibinfo {volume} {9}},\ \bibinfo
  {pages} {353} (\bibinfo {year} {1922}{\natexlab{a}})}\BibitemShut {NoStop}%
\bibitem [{\citenamefont {Gerlach}\ and\ \citenamefont
  {Stern}(1922{\natexlab{b}})}]{gerlach1922b}%
  \BibitemOpen
  \bibfield  {author} {\bibinfo {author} {\bibfnamefont {W.}~\bibnamefont
  {Gerlach}}\ and\ \bibinfo {author} {\bibfnamefont {O.}~\bibnamefont
  {Stern}},\ }\href@noop {} {\bibfield  {journal} {\bibinfo  {journal}
  {Zeitschrift f{\"u}r Physik}\ }\textbf {\bibinfo {volume} {8}},\ \bibinfo
  {pages} {110} (\bibinfo {year} {1922}{\natexlab{b}})}\BibitemShut {NoStop}%
\bibitem [{\citenamefont {Fraser}(1927)}]{Fraser1927}%
  \BibitemOpen
  \bibfield  {author} {\bibinfo {author} {\bibfnamefont {R.~G.}\ \bibnamefont
  {Fraser}},\ }\href@noop {} {\bibfield  {journal} {\bibinfo  {journal}
  {Proceedings of the Royal Society of London. Series A, Containing Papers of a
  Mathematical and Physical Character}\ }\textbf {\bibinfo {volume} {114}},\
  \bibinfo {pages} {212} (\bibinfo {year} {1927})}\BibitemShut {NoStop}%
\bibitem [{\citenamefont {Brillouin}(1927)}]{brillouin1927}%
  \BibitemOpen
  \bibfield  {author} {\bibinfo {author} {\bibfnamefont {L.}~\bibnamefont
  {Brillouin}},\ }\href@noop {} {\bibfield  {journal} {\bibinfo  {journal}
  {CR}\ }\textbf {\bibinfo {volume} {184}},\ \bibinfo {pages} {82} (\bibinfo
  {year} {1927})}\BibitemShut {NoStop}%
\bibitem [{\citenamefont {Brillouin}(1928)}]{brillouin1928}%
  \BibitemOpen
  \bibfield  {author} {\bibinfo {author} {\bibfnamefont {L.}~\bibnamefont
  {Brillouin}},\ }\href@noop {} {\bibfield  {journal} {\bibinfo  {journal}
  {Proceedings of the National Academy of Sciences}\ }\textbf {\bibinfo
  {volume} {14}},\ \bibinfo {pages} {755} (\bibinfo {year} {1928})}\BibitemShut
  {NoStop}%
\bibitem [{\citenamefont {Garraway}\ and\ \citenamefont
  {Stenholm}(2002)}]{Garraway2002}%
  \BibitemOpen
  \bibfield  {author} {\bibinfo {author} {\bibfnamefont {B.}~\bibnamefont
  {Garraway}}\ and\ \bibinfo {author} {\bibfnamefont {S.}~\bibnamefont
  {Stenholm}},\ }\href@noop {} {\bibfield  {journal} {\bibinfo  {journal}
  {Contemporary Physics}\ }\textbf {\bibinfo {volume} {43}},\ \bibinfo {pages}
  {147} (\bibinfo {year} {2002})}\BibitemShut {NoStop}%
\bibitem [{\citenamefont {Garraway}\ and\ \citenamefont
  {Stenholm}(1999)}]{Garraway1999}%
  \BibitemOpen
  \bibfield  {author} {\bibinfo {author} {\bibfnamefont {B.}~\bibnamefont
  {Garraway}}\ and\ \bibinfo {author} {\bibfnamefont {S.}~\bibnamefont
  {Stenholm}},\ }\href@noop {} {\bibfield  {journal} {\bibinfo  {journal}
  {Physical Review A}\ }\textbf {\bibinfo {volume} {60}},\ \bibinfo {pages}
  {63} (\bibinfo {year} {1999})}\BibitemShut {NoStop}%
\bibitem [{\citenamefont {Gallup}\ \emph {et~al.}(2001)\citenamefont {Gallup},
  \citenamefont {Batelaan},\ and\ \citenamefont {Gay}}]{Gallup2001}%
  \BibitemOpen
  \bibfield  {author} {\bibinfo {author} {\bibfnamefont {G.~A.}\ \bibnamefont
  {Gallup}}, \bibinfo {author} {\bibfnamefont {H.}~\bibnamefont {Batelaan}}, \
  and\ \bibinfo {author} {\bibfnamefont {T.~J.}\ \bibnamefont {Gay}},\ }\href
  {\doibase 10.1103/PhysRevLett.86.4508} {\bibfield  {journal} {\bibinfo
  {journal} {Phys. Rev. Lett.}\ }\textbf {\bibinfo {volume} {86}},\ \bibinfo
  {pages} {4508} (\bibinfo {year} {2001})}\BibitemShut {NoStop}%
\bibitem [{\citenamefont {Mott}(1929)}]{Mott1929}%
  \BibitemOpen
  \bibfield  {author} {\bibinfo {author} {\bibfnamefont {N.~F.}\ \bibnamefont
  {Mott}},\ }\href@noop {} {\bibfield  {journal} {\bibinfo  {journal}
  {Proceedings of the Royal Society of London. Series A, Containing Papers of a
  Mathematical and Physical Character}\ }\textbf {\bibinfo {volume} {124}},\
  \bibinfo {pages} {425} (\bibinfo {year} {1929})}\BibitemShut {NoStop}%
\bibitem [{\citenamefont {Kessler}(1985)}]{Kessler1985}%
  \BibitemOpen
  \bibfield  {author} {\bibinfo {author} {\bibfnamefont {J.}~\bibnamefont
  {Kessler}},\ }\href@noop {} {\emph {\bibinfo {title} {Polarized
  electrons}}},\ Vol.~\bibinfo {volume} {1}\ (\bibinfo  {publisher} {Springer
  Science \& Business Media},\ \bibinfo {year} {1985})\BibitemShut {NoStop}%
\bibitem [{\citenamefont {Fabian}\ \emph {et~al.}(2007)\citenamefont {Fabian},
  \citenamefont {Abiague}, \citenamefont {Ertler}, \citenamefont {Stano},\ and\
  \citenamefont {Zutic}}]{fabian2007}%
  \BibitemOpen
  \bibfield  {author} {\bibinfo {author} {\bibfnamefont {J.}~\bibnamefont
  {Fabian}}, \bibinfo {author} {\bibfnamefont {A.}~\bibnamefont {Abiague}},
  \bibinfo {author} {\bibfnamefont {C.}~\bibnamefont {Ertler}}, \bibinfo
  {author} {\bibfnamefont {P.}~\bibnamefont {Stano}}, \ and\ \bibinfo {author}
  {\bibfnamefont {I.}~\bibnamefont {Zutic}},\ }\href@noop {} {\bibfield
  {journal} {\bibinfo  {journal} {Acta Physica Slovaca}\ }\textbf {\bibinfo
  {volume} {57}} (\bibinfo {year} {2007})}\BibitemShut {NoStop}%
\bibitem [{\citenamefont {Fabian}\ \emph {et~al.}(2002)\citenamefont {Fabian},
  \citenamefont {\ifmmode \check{Z}\else \v{Z}\fi{}uti\ifmmode~\acute{c}\else
  \'{c}\fi{}},\ and\ \citenamefont {Sarma}}]{Fabian2002}%
  \BibitemOpen
  \bibfield  {author} {\bibinfo {author} {\bibfnamefont {J.}~\bibnamefont
  {Fabian}}, \bibinfo {author} {\bibfnamefont {I.}~\bibnamefont {\ifmmode
  \check{Z}\else \v{Z}\fi{}uti\ifmmode~\acute{c}\else \'{c}\fi{}}}, \ and\
  \bibinfo {author} {\bibfnamefont {S.~D.}\ \bibnamefont {Sarma}},\ }\href
  {\doibase 10.1103/PhysRevB.66.165301} {\bibfield  {journal} {\bibinfo
  {journal} {Phys. Rev. B}\ }\textbf {\bibinfo {volume} {66}},\ \bibinfo
  {pages} {165301} (\bibinfo {year} {2002})}\BibitemShut {NoStop}%
\bibitem [{\citenamefont {Martin}(2003)}]{Martin2003}%
  \BibitemOpen
  \bibfield  {author} {\bibinfo {author} {\bibfnamefont {I.}~\bibnamefont
  {Martin}},\ }\href {\doibase 10.1103/PhysRevB.67.014421} {\bibfield
  {journal} {\bibinfo  {journal} {Physical Review B}\ }\textbf {\bibinfo
  {volume} {67}},\ \bibinfo {pages} {014421} (\bibinfo {year} {2003})},\
  \Eprint {http://arxiv.org/abs/0201481} {arXiv:0201481} \BibitemShut {NoStop}%
\bibitem [{\citenamefont {Wr{\'o}bel}\ \emph {et~al.}(2004)\citenamefont
  {Wr{\'o}bel}, \citenamefont {Dietl}, \citenamefont {{\L}usakowski},
  \citenamefont {Grabecki}, \citenamefont {Fronc}, \citenamefont {Hey},
  \citenamefont {Ploog},\ and\ \citenamefont {Shtrikman}}]{Wrobel2004}%
  \BibitemOpen
  \bibfield  {author} {\bibinfo {author} {\bibfnamefont {J.}~\bibnamefont
  {Wr{\'o}bel}}, \bibinfo {author} {\bibfnamefont {T.}~\bibnamefont {Dietl}},
  \bibinfo {author} {\bibfnamefont {A.}~\bibnamefont {{\L}usakowski}}, \bibinfo
  {author} {\bibfnamefont {G.}~\bibnamefont {Grabecki}}, \bibinfo {author}
  {\bibfnamefont {K.}~\bibnamefont {Fronc}}, \bibinfo {author} {\bibfnamefont
  {R.}~\bibnamefont {Hey}}, \bibinfo {author} {\bibfnamefont {K.}~\bibnamefont
  {Ploog}}, \ and\ \bibinfo {author} {\bibfnamefont {H.}~\bibnamefont
  {Shtrikman}},\ }\href@noop {} {\bibfield  {journal} {\bibinfo  {journal}
  {Physical review letters}\ }\textbf {\bibinfo {volume} {93}},\ \bibinfo
  {pages} {246601} (\bibinfo {year} {2004})}\BibitemShut {NoStop}%
\bibitem [{\citenamefont {Fabian}\ and\ \citenamefont
  {Sarma}(2002)}]{Fabian2002b}%
  \BibitemOpen
  \bibfield  {author} {\bibinfo {author} {\bibfnamefont {J.}~\bibnamefont
  {Fabian}}\ and\ \bibinfo {author} {\bibfnamefont {S.~D.}\ \bibnamefont
  {Sarma}},\ }\href@noop {} {\bibfield  {journal} {\bibinfo  {journal}
  {Physical Review B}\ }\textbf {\bibinfo {volume} {66}},\ \bibinfo {pages}
  {024436} (\bibinfo {year} {2002})}\BibitemShut {NoStop}%
\bibitem [{\citenamefont {D'yakonov}\ and\ \citenamefont
  {Perel'}(1971)}]{Dyakonov1971}%
  \BibitemOpen
  \bibfield  {author} {\bibinfo {author} {\bibfnamefont {M.~I.}\ \bibnamefont
  {D'yakonov}}\ and\ \bibinfo {author} {\bibfnamefont {V.~I.}\ \bibnamefont
  {Perel'}},\ }\href@noop {} {\bibfield  {journal} {\bibinfo  {journal}
  {Physics Letters A}\ }\textbf {\bibinfo {volume} {35}},\ \bibinfo {pages}
  {459} (\bibinfo {year} {1971})}\BibitemShut {NoStop}%
\bibitem [{\citenamefont {{D'Yakonov}}\ and\ \citenamefont
  {{Perel'}}(1971)}]{Dyakonov1971a}%
  \BibitemOpen
  \bibfield  {author} {\bibinfo {author} {\bibfnamefont {M.~I.}\ \bibnamefont
  {{D'Yakonov}}}\ and\ \bibinfo {author} {\bibfnamefont {V.~I.}\ \bibnamefont
  {{Perel'}}},\ }\href@noop {} {\bibfield  {journal} {\bibinfo  {journal}
  {ZhETF Pisma Redaktsiiu}\ }\textbf {\bibinfo {volume} {13}},\ \bibinfo
  {pages} {657} (\bibinfo {year} {1971})}\BibitemShut {NoStop}%
\bibitem [{\citenamefont {Nagaosa}\ \emph {et~al.}(2010)\citenamefont
  {Nagaosa}, \citenamefont {Sinova}, \citenamefont {Onoda}, \citenamefont
  {MacDonald},\ and\ \citenamefont {Ong}}]{nagaosa2010}%
  \BibitemOpen
  \bibfield  {author} {\bibinfo {author} {\bibfnamefont {N.}~\bibnamefont
  {Nagaosa}}, \bibinfo {author} {\bibfnamefont {J.}~\bibnamefont {Sinova}},
  \bibinfo {author} {\bibfnamefont {S.}~\bibnamefont {Onoda}}, \bibinfo
  {author} {\bibfnamefont {A.~H.}\ \bibnamefont {MacDonald}}, \ and\ \bibinfo
  {author} {\bibfnamefont {N.~P.}\ \bibnamefont {Ong}},\ }\href@noop {}
  {\bibfield  {journal} {\bibinfo  {journal} {Reviews of modern physics}\
  }\textbf {\bibinfo {volume} {82}},\ \bibinfo {pages} {1539} (\bibinfo {year}
  {2010})}\BibitemShut {NoStop}%
\bibitem [{\citenamefont {Sinova}\ \emph {et~al.}(2015)\citenamefont {Sinova},
  \citenamefont {Valenzuela}, \citenamefont {Wunderlich}, \citenamefont
  {Back},\ and\ \citenamefont {Jungwirth}}]{Sinova2015}%
  \BibitemOpen
  \bibfield  {author} {\bibinfo {author} {\bibfnamefont {J.}~\bibnamefont
  {Sinova}}, \bibinfo {author} {\bibfnamefont {S.~O.}\ \bibnamefont
  {Valenzuela}}, \bibinfo {author} {\bibfnamefont {J.}~\bibnamefont
  {Wunderlich}}, \bibinfo {author} {\bibfnamefont {C.}~\bibnamefont {Back}}, \
  and\ \bibinfo {author} {\bibfnamefont {T.}~\bibnamefont {Jungwirth}},\
  }\href@noop {} {\bibfield  {journal} {\bibinfo  {journal} {Reviews of modern
  physics}\ }\textbf {\bibinfo {volume} {87}},\ \bibinfo {pages} {1213}
  (\bibinfo {year} {2015})}\BibitemShut {NoStop}%
\bibitem [{\citenamefont {Edelstein}(1990)}]{Edelstein1990}%
  \BibitemOpen
  \bibfield  {author} {\bibinfo {author} {\bibfnamefont {V.~M.}\ \bibnamefont
  {Edelstein}},\ }\href@noop {} {\bibfield  {journal} {\bibinfo  {journal}
  {Solid State Communications}\ }\textbf {\bibinfo {volume} {73}},\ \bibinfo
  {pages} {233} (\bibinfo {year} {1990})}\BibitemShut {NoStop}%
\bibitem [{\citenamefont {Ganichev}\ \emph {et~al.}(2002)\citenamefont
  {Ganichev}, \citenamefont {Ivchenko},\ and\ \citenamefont
  {Belkov}}]{Ganichev2002}%
  \BibitemOpen
  \bibfield  {author} {\bibinfo {author} {\bibfnamefont {S.}~\bibnamefont
  {Ganichev}}, \bibinfo {author} {\bibfnamefont {E.}~\bibnamefont {Ivchenko}},
  \ and\ \bibinfo {author} {\bibfnamefont {V.}~\bibnamefont {Belkov}},\
  }\href@noop {} {\bibfield  {journal} {\bibinfo  {journal} {Nature}\ }\textbf
  {\bibinfo {volume} {417}},\ \bibinfo {pages} {153} (\bibinfo {year}
  {2002})}\BibitemShut {NoStop}%
\bibitem [{\citenamefont {Kato}\ \emph {et~al.}(2004)\citenamefont {Kato},
  \citenamefont {Myers}, \citenamefont {Gossard},\ and\ \citenamefont
  {Awschalom}}]{Kato2004c}%
  \BibitemOpen
  \bibfield  {author} {\bibinfo {author} {\bibfnamefont {Y.~K.}\ \bibnamefont
  {Kato}}, \bibinfo {author} {\bibfnamefont {R.~C.}\ \bibnamefont {Myers}},
  \bibinfo {author} {\bibfnamefont {A.~C.}\ \bibnamefont {Gossard}}, \ and\
  \bibinfo {author} {\bibfnamefont {D.~D.}\ \bibnamefont {Awschalom}},\
  }\href@noop {} {\bibfield  {journal} {\bibinfo  {journal} {Nature}\ }\textbf
  {\bibinfo {volume} {427}},\ \bibinfo {pages} {50} (\bibinfo {year}
  {2004})}\BibitemShut {NoStop}%
\bibitem [{\citenamefont {Van~Dam}(1938)}]{vandam1938}%
  \BibitemOpen
  \bibfield  {author} {\bibinfo {author} {\bibfnamefont {L.}~\bibnamefont
  {Van~Dam}},\ }\emph {\bibinfo {title} {On the utilization of oxygen and
  regulation of breathing in some aquatic animals}},\ \href@noop {} {Ph.D.
  thesis},\ \bibinfo  {school} {Drukkerij Volharding} (\bibinfo {year}
  {1938})\BibitemShut {NoStop}%
\bibitem [{\citenamefont {Hazelhoff}\ and\ \citenamefont
  {Evenhuis}(1952)}]{Hazelhoff1952}%
  \BibitemOpen
  \bibfield  {author} {\bibinfo {author} {\bibfnamefont {E.~H.}\ \bibnamefont
  {Hazelhoff}}\ and\ \bibinfo {author} {\bibfnamefont {H.~H.}\ \bibnamefont
  {Evenhuis}},\ }\href {\doibase 10.1038/169077a0} {\bibfield  {journal}
  {\bibinfo  {journal} {Nature}\ }\textbf {\bibinfo {volume} {169}},\ \bibinfo
  {pages} {77} (\bibinfo {year} {1952})}\BibitemShut {NoStop}%
\bibitem [{\citenamefont {Hughes}\ and\ \citenamefont
  {Morgan}(1973)}]{Hughes1973}%
  \BibitemOpen
  \bibfield  {author} {\bibinfo {author} {\bibfnamefont {G.~M.}\ \bibnamefont
  {Hughes}}\ and\ \bibinfo {author} {\bibfnamefont {M.}~\bibnamefont
  {Morgan}},\ }\href {\doibase 10.1111/j.1469-185X.1973.tb01009.x} {\bibfield
  {journal} {\bibinfo  {journal} {Biological Reviews}\ }\textbf {\bibinfo
  {volume} {48}},\ \bibinfo {pages} {419} (\bibinfo {year} {1973})}\BibitemShut
  {NoStop}%
\bibitem [{\citenamefont {Ullrich}\ \emph {et~al.}(1961)\citenamefont
  {Ullrich}, \citenamefont {Kramer},\ and\ \citenamefont
  {Boylan}}]{Ullrich1961}%
  \BibitemOpen
  \bibfield  {author} {\bibinfo {author} {\bibfnamefont {K.~J.}\ \bibnamefont
  {Ullrich}}, \bibinfo {author} {\bibfnamefont {K.}~\bibnamefont {Kramer}}, \
  and\ \bibinfo {author} {\bibfnamefont {J.~W.}\ \bibnamefont {Boylan}},\
  }\href@noop {} {\bibfield  {journal} {\bibinfo  {journal} {Progress in
  cardiovascular diseases}\ }\textbf {\bibinfo {volume} {3}},\ \bibinfo {pages}
  {395} (\bibinfo {year} {1961})}\BibitemShut {NoStop}%
\bibitem [{\citenamefont {Randall}\ and\ \citenamefont
  {Longtin}(1939)}]{Randall1939}%
  \BibitemOpen
  \bibfield  {author} {\bibinfo {author} {\bibfnamefont {M.}~\bibnamefont
  {Randall}}\ and\ \bibinfo {author} {\bibfnamefont {B.}~\bibnamefont
  {Longtin}},\ }\href {\doibase 10.1021/ie50358a032} {\bibfield  {journal}
  {\bibinfo  {journal} {Industrial \& Engineering Chemistry}\ }\textbf
  {\bibinfo {volume} {31}},\ \bibinfo {pages} {1295} (\bibinfo {year}
  {1939})}\BibitemShut {NoStop}%
\bibitem [{\citenamefont {Schicks}\ \emph {et~al.}(2019)\citenamefont
  {Schicks}, \citenamefont {Spangenberg}, \citenamefont {Giese}, \citenamefont
  {Luzi-Helbing}, \citenamefont {Priegnitz}, \citenamefont {Heeschen},
  \citenamefont {Strauch}, \citenamefont {Schr{\"o}tter}, \citenamefont
  {K{\"u}ck}, \citenamefont {T{\"o}pfer} \emph {et~al.}}]{Schicks2019}%
  \BibitemOpen
  \bibfield  {author} {\bibinfo {author} {\bibfnamefont {J.~M.}\ \bibnamefont
  {Schicks}}, \bibinfo {author} {\bibfnamefont {E.}~\bibnamefont
  {Spangenberg}}, \bibinfo {author} {\bibfnamefont {R.}~\bibnamefont {Giese}},
  \bibinfo {author} {\bibfnamefont {M.}~\bibnamefont {Luzi-Helbing}}, \bibinfo
  {author} {\bibfnamefont {M.}~\bibnamefont {Priegnitz}}, \bibinfo {author}
  {\bibfnamefont {K.~U.}\ \bibnamefont {Heeschen}}, \bibinfo {author}
  {\bibfnamefont {B.}~\bibnamefont {Strauch}}, \bibinfo {author} {\bibfnamefont
  {J.}~\bibnamefont {Schr{\"o}tter}}, \bibinfo {author} {\bibfnamefont
  {J.}~\bibnamefont {K{\"u}ck}}, \bibinfo {author} {\bibfnamefont
  {M.}~\bibnamefont {T{\"o}pfer}},  \emph {et~al.},\ }in\ \href@noop {} {\emph
  {\bibinfo {booktitle} {Offshore Technology Conference}}}\ (\bibinfo
  {organization} {OnePetro},\ \bibinfo {year} {2019})\BibitemShut {NoStop}%
\bibitem [{\citenamefont {Kos}\ \emph {et~al.}(2007)\citenamefont {Kos},
  \citenamefont {Hruska}, \citenamefont {Crooker}, \citenamefont {Saxena},\
  and\ \citenamefont {Smith}}]{kos2007}%
  \BibitemOpen
  \bibfield  {author} {\bibinfo {author} {\bibfnamefont {S.}~\bibnamefont
  {Kos}}, \bibinfo {author} {\bibfnamefont {M.}~\bibnamefont {Hruska}},
  \bibinfo {author} {\bibfnamefont {S.~A.}\ \bibnamefont {Crooker}}, \bibinfo
  {author} {\bibfnamefont {A.}~\bibnamefont {Saxena}}, \ and\ \bibinfo {author}
  {\bibfnamefont {D.~L.}\ \bibnamefont {Smith}},\ }\href@noop {} {\bibfield
  {journal} {\bibinfo  {journal} {Computing in Science \& Engineering}\
  }\textbf {\bibinfo {volume} {9}},\ \bibinfo {pages} {46} (\bibinfo {year}
  {2007})}\BibitemShut {NoStop}%
\bibitem [{\citenamefont {Wu}\ \emph {et~al.}(2010)\citenamefont {Wu},
  \citenamefont {Jiang},\ and\ \citenamefont {Weng}}]{Wu2010}%
  \BibitemOpen
  \bibfield  {author} {\bibinfo {author} {\bibfnamefont {M.}~\bibnamefont
  {Wu}}, \bibinfo {author} {\bibfnamefont {J.}~\bibnamefont {Jiang}}, \ and\
  \bibinfo {author} {\bibfnamefont {M.}~\bibnamefont {Weng}},\ }\href@noop {}
  {\bibfield  {journal} {\bibinfo  {journal} {Physics Reports}\ }\textbf
  {\bibinfo {volume} {493}},\ \bibinfo {pages} {61} (\bibinfo {year}
  {2010})}\BibitemShut {NoStop}%
\bibitem [{\citenamefont {Harmon}\ \emph {et~al.}(2015)\citenamefont {Harmon},
  \citenamefont {Peterson}, \citenamefont {Geppert}, \citenamefont {Patel},
  \citenamefont {Palmstr{\o}m}, \citenamefont {Crowell},\ and\ \citenamefont
  {Flatt{\'{e}}}}]{Harmon2015}%
  \BibitemOpen
  \bibfield  {author} {\bibinfo {author} {\bibfnamefont {N.~J.}\ \bibnamefont
  {Harmon}}, \bibinfo {author} {\bibfnamefont {T.~A.}\ \bibnamefont
  {Peterson}}, \bibinfo {author} {\bibfnamefont {C.~C.}\ \bibnamefont
  {Geppert}}, \bibinfo {author} {\bibfnamefont {S.~J.}\ \bibnamefont {Patel}},
  \bibinfo {author} {\bibfnamefont {C.~J.}\ \bibnamefont {Palmstr{\o}m}},
  \bibinfo {author} {\bibfnamefont {P.~A.}\ \bibnamefont {Crowell}}, \ and\
  \bibinfo {author} {\bibfnamefont {M.~E.}\ \bibnamefont {Flatt{\'{e}}}},\
  }\href {\doibase 10.1103/PhysRevB.92.140201} {\bibfield  {journal} {\bibinfo
  {journal} {Physical Review B}\ }\textbf {\bibinfo {volume} {92}},\ \bibinfo
  {pages} {140201} (\bibinfo {year} {2015})}\BibitemShut {NoStop}%
\bibitem [{\citenamefont {O’Brien}\ \emph {et~al.}(2016)\citenamefont
  {O’Brien}, \citenamefont {Spivak}, \citenamefont {Krueger}, \citenamefont
  {Peterson}, \citenamefont {Erickson}, \citenamefont {Bolon}, \citenamefont
  {Geppert}, \citenamefont {Leighton},\ and\ \citenamefont
  {Crowell}}]{obrien2016}%
  \BibitemOpen
  \bibfield  {author} {\bibinfo {author} {\bibfnamefont {L.}~\bibnamefont
  {O’Brien}}, \bibinfo {author} {\bibfnamefont {D.}~\bibnamefont {Spivak}},
  \bibinfo {author} {\bibfnamefont {N.}~\bibnamefont {Krueger}}, \bibinfo
  {author} {\bibfnamefont {T.}~\bibnamefont {Peterson}}, \bibinfo {author}
  {\bibfnamefont {M.}~\bibnamefont {Erickson}}, \bibinfo {author}
  {\bibfnamefont {B.}~\bibnamefont {Bolon}}, \bibinfo {author} {\bibfnamefont
  {C.}~\bibnamefont {Geppert}}, \bibinfo {author} {\bibfnamefont
  {C.}~\bibnamefont {Leighton}}, \ and\ \bibinfo {author} {\bibfnamefont
  {P.}~\bibnamefont {Crowell}},\ }\href@noop {} {\bibfield  {journal} {\bibinfo
   {journal} {Physical Review B}\ }\textbf {\bibinfo {volume} {94}},\ \bibinfo
  {pages} {094431} (\bibinfo {year} {2016})}\BibitemShut {NoStop}%
\bibitem [{\citenamefont {Meier}\ and\ \citenamefont
  {Zachachrenya}(1984)}]{Meier1984}%
  \BibitemOpen
  \bibfield  {author} {\bibinfo {author} {\bibfnamefont {F.}~\bibnamefont
  {Meier}}\ and\ \bibinfo {author} {\bibfnamefont {B.~P.}\ \bibnamefont
  {Zachachrenya}},\ }\href@noop {} {\emph {\bibinfo {title} {Optical
  Orientation: Modern Problems in Condensed Matter Science}}},\ Vol.~\bibinfo
  {volume} {8}\ (\bibinfo  {publisher} {North-Holland},\ \bibinfo {address}
  {Amsterdam},\ \bibinfo {year} {1984})\BibitemShut {NoStop}%
\bibitem [{\citenamefont {Kikkawa}\ and\ \citenamefont
  {Awschalom}(1998)}]{Kikkawa1998}%
  \BibitemOpen
  \bibfield  {author} {\bibinfo {author} {\bibfnamefont {J.~M.}\ \bibnamefont
  {Kikkawa}}\ and\ \bibinfo {author} {\bibfnamefont {D.~D.}\ \bibnamefont
  {Awschalom}},\ }\href@noop {} {\bibfield  {journal} {\bibinfo  {journal}
  {\prl}\ }\textbf {\bibinfo {volume} {80}},\ \bibinfo {pages} {4313} (\bibinfo
  {year} {1998})}\BibitemShut {NoStop}%
\bibitem [{\citenamefont {see Supplemental~Information}()}]{supp}%
  \BibitemOpen
  \bibfield  {author} {\bibinfo {author} {\bibnamefont {see
  Supplemental~Information}},\ }\href@noop {} {}\BibitemShut {NoStop}%
\bibitem [{\citenamefont {Kikkawa}\ \emph {et~al.}(1997)\citenamefont
  {Kikkawa}, \citenamefont {Smorchkova}, \citenamefont {Samarth},\ and\
  \citenamefont {Awschalom}}]{Kikkawa1997}%
  \BibitemOpen
  \bibfield  {author} {\bibinfo {author} {\bibfnamefont {J.~M.}\ \bibnamefont
  {Kikkawa}}, \bibinfo {author} {\bibfnamefont {I.~P.}\ \bibnamefont
  {Smorchkova}}, \bibinfo {author} {\bibfnamefont {N.}~\bibnamefont {Samarth}},
  \ and\ \bibinfo {author} {\bibfnamefont {D.~D.}\ \bibnamefont {Awschalom}},\
  }\href@noop {} {\bibfield  {journal} {\bibinfo  {journal} {Science}\ }\textbf
  {\bibinfo {volume} {277}},\ \bibinfo {pages} {1284} (\bibinfo {year}
  {1997})}\BibitemShut {NoStop}%
\bibitem [{\citenamefont {Yu}\ and\ \citenamefont {Flatt\'e}(2002)}]{Yu2002a}%
  \BibitemOpen
  \bibfield  {author} {\bibinfo {author} {\bibfnamefont {Z.~G.}\ \bibnamefont
  {Yu}}\ and\ \bibinfo {author} {\bibfnamefont {M.~E.}\ \bibnamefont
  {Flatt\'e}},\ }\href {\doibase 10.1103/PhysRevB.66.201202} {\bibfield
  {journal} {\bibinfo  {journal} {Phys. Rev. B}\ }\textbf {\bibinfo {volume}
  {66}},\ \bibinfo {pages} {201202(R)} (\bibinfo {year} {2002})}\BibitemShut
  {NoStop}%
\bibitem [{\citenamefont {Yu}\ and\ \citenamefont
  {Flatt{\'{e}}}(2002)}]{Yu2002}%
  \BibitemOpen
  \bibfield  {author} {\bibinfo {author} {\bibfnamefont {Z.~G.}\ \bibnamefont
  {Yu}}\ and\ \bibinfo {author} {\bibfnamefont {M.~E.}\ \bibnamefont
  {Flatt{\'{e}}}},\ }\href {\doibase 10.1103/PhysRevB.66.235302} {\bibfield
  {journal} {\bibinfo  {journal} {Physical Review B}\ }\textbf {\bibinfo
  {volume} {66}},\ \bibinfo {pages} {235302} (\bibinfo {year} {2002})},\
  \Eprint {http://arxiv.org/abs/0206321} {0206321} \BibitemShut {NoStop}%
\bibitem [{\citenamefont {Tahara}\ \emph {et~al.}(2016)\citenamefont {Tahara},
  \citenamefont {Ando}, \citenamefont {Kameno}, \citenamefont {Koike},
  \citenamefont {Tanaka}, \citenamefont {Miwa}, \citenamefont {Suzuki},
  \citenamefont {Sasaki}, \citenamefont {Oikawa},\ and\ \citenamefont
  {Shiraishi}}]{tahara2016}%
  \BibitemOpen
  \bibfield  {author} {\bibinfo {author} {\bibfnamefont {T.}~\bibnamefont
  {Tahara}}, \bibinfo {author} {\bibfnamefont {Y.}~\bibnamefont {Ando}},
  \bibinfo {author} {\bibfnamefont {M.}~\bibnamefont {Kameno}}, \bibinfo
  {author} {\bibfnamefont {H.}~\bibnamefont {Koike}}, \bibinfo {author}
  {\bibfnamefont {K.}~\bibnamefont {Tanaka}}, \bibinfo {author} {\bibfnamefont
  {S.}~\bibnamefont {Miwa}}, \bibinfo {author} {\bibfnamefont {Y.}~\bibnamefont
  {Suzuki}}, \bibinfo {author} {\bibfnamefont {T.}~\bibnamefont {Sasaki}},
  \bibinfo {author} {\bibfnamefont {T.}~\bibnamefont {Oikawa}}, \ and\ \bibinfo
  {author} {\bibfnamefont {M.}~\bibnamefont {Shiraishi}},\ }\href@noop {}
  {\bibfield  {journal} {\bibinfo  {journal} {Physical Review B}\ }\textbf
  {\bibinfo {volume} {93}},\ \bibinfo {pages} {214406} (\bibinfo {year}
  {2016})}\BibitemShut {NoStop}%
\bibitem [{\citenamefont {Ohno}\ \emph {et~al.}(1999)\citenamefont {Ohno},
  \citenamefont {Terauchi}, \citenamefont {Adachi}, \citenamefont {Matsukura},\
  and\ \citenamefont {Ohno}}]{Ohno1999b}%
  \BibitemOpen
  \bibfield  {author} {\bibinfo {author} {\bibfnamefont {Y.}~\bibnamefont
  {Ohno}}, \bibinfo {author} {\bibfnamefont {R.}~\bibnamefont {Terauchi}},
  \bibinfo {author} {\bibfnamefont {T.}~\bibnamefont {Adachi}}, \bibinfo
  {author} {\bibfnamefont {F.}~\bibnamefont {Matsukura}}, \ and\ \bibinfo
  {author} {\bibfnamefont {H.}~\bibnamefont {Ohno}},\ }\href@noop {} {\bibfield
   {journal} {\bibinfo  {journal} {\prl}\ }\textbf {\bibinfo {volume} {83}},\
  \bibinfo {pages} {4196} (\bibinfo {year} {1999})}\BibitemShut {NoStop}%
\bibitem [{\citenamefont {Lou}\ \emph {et~al.}(2007)\citenamefont {Lou},
  \citenamefont {Adelmann}, \citenamefont {Crooker}, \citenamefont {Garlid},
  \citenamefont {Zhang}, \citenamefont {Reddy}, \citenamefont {Flexner},
  \citenamefont {Palmst{r\o m}},\ and\ \citenamefont {Crowell}}]{Lou2007}%
  \BibitemOpen
  \bibfield  {author} {\bibinfo {author} {\bibfnamefont {X.}~\bibnamefont
  {Lou}}, \bibinfo {author} {\bibfnamefont {C.}~\bibnamefont {Adelmann}},
  \bibinfo {author} {\bibfnamefont {S.~A.}\ \bibnamefont {Crooker}}, \bibinfo
  {author} {\bibfnamefont {E.~S.}\ \bibnamefont {Garlid}}, \bibinfo {author}
  {\bibfnamefont {J.}~\bibnamefont {Zhang}}, \bibinfo {author} {\bibfnamefont
  {K.~S.~M.}\ \bibnamefont {Reddy}}, \bibinfo {author} {\bibfnamefont {S.~D.}\
  \bibnamefont {Flexner}}, \bibinfo {author} {\bibfnamefont {C.~J.}\
  \bibnamefont {Palmst{r\o m}}}, \ and\ \bibinfo {author} {\bibfnamefont
  {P.~A.}\ \bibnamefont {Crowell}},\ }\href {\doibase 10.1038/nphys543}
  {\bibfield  {journal} {\bibinfo  {journal} {Nature Physics}\ }\textbf
  {\bibinfo {volume} {3}},\ \bibinfo {pages} {197} (\bibinfo {year}
  {2007})}\BibitemShut {NoStop}%
\bibitem [{\citenamefont {Furis}\ \emph {et~al.}(2007)\citenamefont {Furis},
  \citenamefont {Smith}, \citenamefont {Kos}, \citenamefont {Garlid},
  \citenamefont {Reddy}, \citenamefont {Palmstr{\o}m}, \citenamefont
  {Crowell},\ and\ \citenamefont {Crooker}}]{Furis2007}%
  \BibitemOpen
  \bibfield  {author} {\bibinfo {author} {\bibfnamefont {M.}~\bibnamefont
  {Furis}}, \bibinfo {author} {\bibfnamefont {D.~L.}\ \bibnamefont {Smith}},
  \bibinfo {author} {\bibfnamefont {S.}~\bibnamefont {Kos}}, \bibinfo {author}
  {\bibfnamefont {E.~S.}\ \bibnamefont {Garlid}}, \bibinfo {author}
  {\bibfnamefont {K.~S.~M.}\ \bibnamefont {Reddy}}, \bibinfo {author}
  {\bibfnamefont {C.~J.}\ \bibnamefont {Palmstr{\o}m}}, \bibinfo {author}
  {\bibfnamefont {P.~a.}\ \bibnamefont {Crowell}}, \ and\ \bibinfo {author}
  {\bibfnamefont {S.~a.}\ \bibnamefont {Crooker}},\ }\href {\doibase
  10.1088/1367-2630/9/9/347} {\bibfield  {journal} {\bibinfo  {journal} {New
  Journal of Physics}\ }\textbf {\bibinfo {volume} {9}},\ \bibinfo {pages}
  {347} (\bibinfo {year} {2007})}\BibitemShut {NoStop}%
\bibitem [{\citenamefont {Olejn{\'\i}k}\ \emph {et~al.}(2012)\citenamefont
  {Olejn{\'\i}k}, \citenamefont {Wunderlich}, \citenamefont {Irvine},
  \citenamefont {Campion}, \citenamefont {Amin}, \citenamefont {Sinova},\ and\
  \citenamefont {Jungwirth}}]{olejnik2012}%
  \BibitemOpen
  \bibfield  {author} {\bibinfo {author} {\bibfnamefont {K.}~\bibnamefont
  {Olejn{\'\i}k}}, \bibinfo {author} {\bibfnamefont {J.}~\bibnamefont
  {Wunderlich}}, \bibinfo {author} {\bibfnamefont {A.}~\bibnamefont {Irvine}},
  \bibinfo {author} {\bibfnamefont {R.}~\bibnamefont {Campion}}, \bibinfo
  {author} {\bibfnamefont {V.}~\bibnamefont {Amin}}, \bibinfo {author}
  {\bibfnamefont {J.}~\bibnamefont {Sinova}}, \ and\ \bibinfo {author}
  {\bibfnamefont {T.}~\bibnamefont {Jungwirth}},\ }\href@noop {} {\bibfield
  {journal} {\bibinfo  {journal} {Physical Review Letters}\ }\textbf {\bibinfo
  {volume} {109}},\ \bibinfo {pages} {076601} (\bibinfo {year}
  {2012})}\BibitemShut {NoStop}%
\bibitem [{\citenamefont {Cohen}(2009)}]{Cohen2009}%
  \BibitemOpen
  \bibfield  {author} {\bibinfo {author} {\bibfnamefont {A.}~\bibnamefont
  {Cohen}},\ }\href@noop {} {\bibfield  {journal} {\bibinfo  {journal} {J.
  Phys. Chem. A}\ }\textbf {\bibinfo {volume} {113}},\ \bibinfo {pages} {11084}
  (\bibinfo {year} {2009})}\BibitemShut {NoStop}%
\bibitem [{\citenamefont {Kustov}\ \emph {et~al.}(2010)\citenamefont {Kustov},
  \citenamefont {Laczkowski}, \citenamefont {Hykel}, \citenamefont
  {Hasselbach}, \citenamefont {Dumas-Bouchiat}, \citenamefont {O’Brien},
  \citenamefont {Kauffmann}, \citenamefont {Grechishkin}, \citenamefont
  {Givord}, \citenamefont {Reyne}, \citenamefont {Cugat},\ and\ \citenamefont
  {Dempsey}}]{kustov2010}%
  \BibitemOpen
  \bibfield  {author} {\bibinfo {author} {\bibfnamefont {M.}~\bibnamefont
  {Kustov}}, \bibinfo {author} {\bibfnamefont {P.}~\bibnamefont {Laczkowski}},
  \bibinfo {author} {\bibfnamefont {D.}~\bibnamefont {Hykel}}, \bibinfo
  {author} {\bibfnamefont {K.}~\bibnamefont {Hasselbach}}, \bibinfo {author}
  {\bibfnamefont {F.}~\bibnamefont {Dumas-Bouchiat}}, \bibinfo {author}
  {\bibfnamefont {D.}~\bibnamefont {O’Brien}}, \bibinfo {author}
  {\bibfnamefont {P.}~\bibnamefont {Kauffmann}}, \bibinfo {author}
  {\bibfnamefont {R.}~\bibnamefont {Grechishkin}}, \bibinfo {author}
  {\bibfnamefont {D.}~\bibnamefont {Givord}}, \bibinfo {author} {\bibfnamefont
  {G.}~\bibnamefont {Reyne}}, \bibinfo {author} {\bibfnamefont
  {O.}~\bibnamefont {Cugat}}, \ and\ \bibinfo {author} {\bibfnamefont {N.~M.}\
  \bibnamefont {Dempsey}},\ }\href {\doibase 10.1063/1.3486513} {\bibfield
  {journal} {\bibinfo  {journal} {Journal of Applied Physics}\ }\textbf
  {\bibinfo {volume} {108}},\ \bibinfo {pages} {063914} (\bibinfo {year}
  {2010})}\BibitemShut {NoStop}%
\bibitem [{\citenamefont {Zablotskii}\ \emph {et~al.}(2016)\citenamefont
  {Zablotskii}, \citenamefont {Polyakova}, \citenamefont {Lunov},\ and\
  \citenamefont {Dejneka}}]{zablotskii2016}%
  \BibitemOpen
  \bibfield  {author} {\bibinfo {author} {\bibfnamefont {V.}~\bibnamefont
  {Zablotskii}}, \bibinfo {author} {\bibfnamefont {T.}~\bibnamefont
  {Polyakova}}, \bibinfo {author} {\bibfnamefont {O.}~\bibnamefont {Lunov}}, \
  and\ \bibinfo {author} {\bibfnamefont {A.}~\bibnamefont {Dejneka}},\ }\href
  {\doibase 10.1038/srep37407} {\bibfield  {journal} {\bibinfo  {journal}
  {Scientific Reports}\ }\textbf {\bibinfo {volume} {6}},\ \bibinfo {pages}
  {37407} (\bibinfo {year} {2016})}\BibitemShut {NoStop}%
\bibitem [{\citenamefont {Kohda}\ \emph {et~al.}(2012)\citenamefont {Kohda},
  \citenamefont {Nakamura}, \citenamefont {Nishihara}, \citenamefont
  {Kobayashi}, \citenamefont {Ono}, \citenamefont {Ohe}, \citenamefont
  {Tokura}, \citenamefont {Mineno},\ and\ \citenamefont {Nitta}}]{Kohda2012}%
  \BibitemOpen
  \bibfield  {author} {\bibinfo {author} {\bibfnamefont {M.}~\bibnamefont
  {Kohda}}, \bibinfo {author} {\bibfnamefont {S.}~\bibnamefont {Nakamura}},
  \bibinfo {author} {\bibfnamefont {Y.}~\bibnamefont {Nishihara}}, \bibinfo
  {author} {\bibfnamefont {K.}~\bibnamefont {Kobayashi}}, \bibinfo {author}
  {\bibfnamefont {T.}~\bibnamefont {Ono}}, \bibinfo {author} {\bibfnamefont
  {J.-i.}\ \bibnamefont {Ohe}}, \bibinfo {author} {\bibfnamefont
  {Y.}~\bibnamefont {Tokura}}, \bibinfo {author} {\bibfnamefont
  {T.}~\bibnamefont {Mineno}}, \ and\ \bibinfo {author} {\bibfnamefont
  {J.}~\bibnamefont {Nitta}},\ }\href@noop {} {\bibfield  {journal} {\bibinfo
  {journal} {Nature communications}\ }\textbf {\bibinfo {volume} {3}},\
  \bibinfo {pages} {1082} (\bibinfo {year} {2012})}\BibitemShut {NoStop}%
\bibitem [{\citenamefont {Paget}\ \emph {et~al.}(1977)\citenamefont {Paget},
  \citenamefont {Lampel}, \citenamefont {Sapoval},\ and\ \citenamefont
  {Safarov}}]{Paget1977}%
  \BibitemOpen
  \bibfield  {author} {\bibinfo {author} {\bibfnamefont {D.}~\bibnamefont
  {Paget}}, \bibinfo {author} {\bibfnamefont {G.}~\bibnamefont {Lampel}},
  \bibinfo {author} {\bibfnamefont {B.}~\bibnamefont {Sapoval}}, \ and\
  \bibinfo {author} {\bibfnamefont {V.~I.}\ \bibnamefont {Safarov}},\ }\href
  {\doibase 10.1103/PhysRevB.15.5780} {\bibfield  {journal} {\bibinfo
  {journal} {Phys. Rev. B}\ }\textbf {\bibinfo {volume} {15}},\ \bibinfo
  {pages} {5780} (\bibinfo {year} {1977})}\BibitemShut {NoStop}%
\bibitem [{\citenamefont {Chan}\ \emph {et~al.}(2009)\citenamefont {Chan},
  \citenamefont {Hu}, \citenamefont {Zhang}, \citenamefont {Kondo},
  \citenamefont {Palmstr{\o}m},\ and\ \citenamefont {Crowell}}]{Chan2009}%
  \BibitemOpen
  \bibfield  {author} {\bibinfo {author} {\bibfnamefont {M.~K.}\ \bibnamefont
  {Chan}}, \bibinfo {author} {\bibfnamefont {Q.~O.}\ \bibnamefont {Hu}},
  \bibinfo {author} {\bibfnamefont {J.}~\bibnamefont {Zhang}}, \bibinfo
  {author} {\bibfnamefont {T.}~\bibnamefont {Kondo}}, \bibinfo {author}
  {\bibfnamefont {C.~J.}\ \bibnamefont {Palmstr{\o}m}}, \ and\ \bibinfo
  {author} {\bibfnamefont {P.~A.}\ \bibnamefont {Crowell}},\ }\href@noop {}
  {\bibfield  {journal} {\bibinfo  {journal} {Phys. Rev. B}\ }\textbf {\bibinfo
  {volume} {80}},\ \bibinfo {pages} {161206(R)} (\bibinfo {year}
  {2009})}\BibitemShut {NoStop}%
\bibitem [{\citenamefont {Harmon}\ and\ \citenamefont
  {Flatté}(2022)}]{Harmon2022a}%
  \BibitemOpen
  \bibfield  {author} {\bibinfo {author} {\bibfnamefont {N.~J.}\ \bibnamefont
  {Harmon}}\ and\ \bibinfo {author} {\bibfnamefont {M.~E.}\ \bibnamefont
  {Flatté}},\ }\href {\doibase 10.1103/PhysRevB.106.054207} {\bibfield
  {journal} {\bibinfo  {journal} {Physical Review B}\ }\textbf {\bibinfo
  {volume} {106}},\ \bibinfo {pages} {054207} (\bibinfo {year}
  {2022})}\BibitemShut {NoStop}%
\bibitem [{\citenamefont {Geppert}\ \emph {et~al.}(2013)\citenamefont
  {Geppert}, \citenamefont {Christie}, \citenamefont {Chan}, \citenamefont
  {Patel}, \citenamefont {Palmstr{\o}~M},\ and\ \citenamefont
  {Crowell}}]{geppert2013}%
  \BibitemOpen
  \bibfield  {author} {\bibinfo {author} {\bibfnamefont {C.}~\bibnamefont
  {Geppert}}, \bibinfo {author} {\bibfnamefont {K.}~\bibnamefont {Christie}},
  \bibinfo {author} {\bibfnamefont {M.}~\bibnamefont {Chan}}, \bibinfo {author}
  {\bibfnamefont {S.}~\bibnamefont {Patel}}, \bibinfo {author} {\bibfnamefont
  {C.}~\bibnamefont {Palmstr{\o}~M}}, \ and\ \bibinfo {author} {\bibfnamefont
  {P.}~\bibnamefont {Crowell}},\ }in\ \href@noop {} {\emph {\bibinfo
  {booktitle} {APS March Meeting Abstracts}}},\ Vol.\ \bibinfo {volume} {2013}\
  (\bibinfo {year} {2013})\ pp.\ \bibinfo {pages} {C18--010}\BibitemShut
  {NoStop}%
\end{thebibliography}
\end{document}